\documentclass{article}

\usepackage{arxiv}

\usepackage[utf8]{inputenc} 
\usepackage[T1]{fontenc}    
\usepackage{hyperref}       
\usepackage{url}            
\usepackage{booktabs}       
\usepackage{amsfonts}       
\usepackage{nicefrac}       
\usepackage{microtype}      
\usepackage{lipsum}
\usepackage[pdftex]{graphicx}
\DeclareGraphicsExtensions{.pdf,.jpeg,.png}
\usepackage[cmex10]{amsmath}

\title{Systematic Enhancement of Functional Connectivity in Brain-Computer Interfacing using Common Spatial Patterns and Tangent Space Mapping}

\author{
  Saugat Bhattacharyya\\
  School of Computer Science and Electronic Engineering\\
  University of Essex\\
  Colchester, CO1 1RQ, UK \\
  \texttt{saugat.bhattacharyya@essex.ac.uk} \\
   \And
  Mitsuhiro Hayashibe \\
  Department of Robotics\\
  Tohoku University\\
  Japan \\
  \texttt{hayashibe@tohoku.ac.jp} \\
}

\begin{document}
\maketitle

\begin{abstract}
Functional connectivity of cognitive tasks allows researchers to analyse the interaction mapping occurring between different regions of the brain using electroencephalography (EEG) signals. Standard practice in functional connectivity involve studying the electrode pair interactions across several trials. As the cognitive task always involves the human factor, it is inevitable to have lower quality data from the brain signals influenced by the subject concentration or other mental states which can occur anytime over the whole experimental trials. The connectivity among electrodes are heavily influenced by these low quality EEG. In this paper, we aim at enhancing the functional connectivity of mental tasks by implementing a classification step in the process to remove those incorrect EEG trials from the available set. The classification step removes the trials which were mis-classified or had a low probability of occurrence to extract only reliable EEG trials. Through our approach, we have successfully improved the separability among graph parameters for different mental tasks. We also observe an improvement in the readability of the connectivity by focusing only on a group of selected channels rather than employing all the channels.
\end{abstract}


\section{Introduction}

The human brain functions as a complex network of connections among its neurons and it remains a challenge for researchers to understand the connectivity patterns for various tasks performed by the brain in day-to-day activity. With the advent of non-invasive imaging techniques, such as electroencephalography (EEG), magnetoencephalography (MEG), functional Magnetic Resonance Imaging (fMRI) and functional Near Infra Red Spectroscopy (fNIRS) \cite{sood2016}, \cite{roy2014}, \cite{bax2016}, there has been a growing interest in studying brain connectivity over the last decade. From available literature, there exist two organizational principles, functional segragation \cite{ton1994} and integration (or connectivity) \cite{tono2003}, that facilitate brain function. Functional connectivity provides information on the coordinated activities of these segragated groups to perform a cognitive task \cite{mahyari2013}.

Functional connectivity is defined in \cite{stam2007} as the statistical dependency between spatially remote neurophysiological events and is the key to understanding how the coordinated and integrated activity of the human brain takes place. To date, linear measures such as correlation \cite{marr2005} and coherence \cite{rocca2014}, \cite{zava2008}, , directed transfer function (DTF) \cite{fall2010}, \cite{ghosh2015} and non-linear measures, such as  phase synchrony \cite{bol2010}, \cite{bol2009}have been widely used to quantify functional connectivity. These measures are widely used to study the pairwise interaction between different regions of the brain but they are unable to provide information on the complex relationship between the function and organization of the brain \cite{mahyari2013}, \cite{ali2012}. To characterize the topological behaviour of these large networks, researchers have begun to investigate into the area of complex networks, especially graph theoretic measures and methods \cite{bull2009}. Now, the bivariate interrelation between two neural populations could be represented as graphs, where the nodes correspond to the individual brain sites and the edges represents the interaction weighed by the functional connectivity \cite{ali2012}.

So far, the primary focus of functional connectivity in brain research is the study of interaction between the different nodes. Bolanos \emph{et al.} in \cite{bol2010} presented the difference of error response and correct response from subjects performing a decision making task using phase synchrony measures. They inferred that clustering coefficient and binary path length measures differ between error and correct responses. There is very little research which implements functional connectivity in classifying the brain state of the individual. Classifying connectivity features would allow its inclusion in Brain-computer interfaces (BCI) controlled devices, such as robot, prosthesis. Zhang et al. in \cite{zhang2012} have used waveform features, DTF features and their combination to recognize Error Related Potential (ErrP) from the EEG signal and they attained the highest classification accuracy of 85\% in comparison to the accuracy obtained while using the individual waveform and DTF features. Another work by la Rocca \emph{et al.} in \cite{rocca2014} employed spectral coherence connectivity on a dataset comprising of eyes-closed (EC) and eyes-open (EO) EEG. He attained a 100\% recognition accuracy when integrating connectivity in the frontal lobe region and 97.5\% when involving the parieto-occipital lobe region. 

Most studies \cite{rocca2014}, \cite{bol2010}, \cite{zhang2012} on functional connectivity analysis employ all the electrode channels during their processing. Since, the brain performs various processes of the human body in a parallel manner, it is difficult to ascertain the connectivity for a specific task in the brain. Real-time BCI applications use signals from specific regions of the brain and do not require the usage of all channels. So, it would be beneficial if the connectivity is generated only for a selected set of channels. For example, in motor imagery tasks, it is known through literature that the primary motor area and supplementary motor area of brain generates the most dominant signals. Thus, channels such as C3, Cz and C4 are usually selected for analysis in most studies \cite{pfurt1997},\cite{leeb2007}, \cite{ghosh2015}. Such preparation improves the processing time of the BCI but it may decrease the accuracy of the system. It may so happen that in certain subjects other electrode channels (as in this example, say, FC3 instead of C3) may provide better result than the standard ones. Thus, it is necessary to select the optimal set of channels from the complete set for better performance of the BCI. In nodal interaction based studies, one analyses the averages between the interaction of the electrode pairs across several trials. However, in reality, human brain reaction to a certain event varies a lot over different trials depending on other brain processes occurring in a complex way. Other than misclassification by the decoder, human brain actually generates clear responses to the event along with unclear responses as well, being enveloped by other mental states running in parallel. This situation leads naturally to inclusion of misleading data (or trials) in the complete data set which may effect the results of the connectivity analysis significantly. 

Hence, in this study, we have aimed at solving the two issues described above which in turn would provide more relevant information on the interaction among electrodes during a cognitive task.  First, we select the optimal set of electrode channels for each subjects and create a new set of spatially filtered projected signals \cite{wang2005}, \cite{arva2011}. Then, we remove the low-information trials (trials which were incorrectly classified or trials with a low probability of occurrence) and keep the high-information trials(correctly classified trials with a high probability of occurence) for connectivity analysis. In summary, we aim at designing a system which selects an optimal set of channels and perform functional connectivity analysis on them for different correctly classified mental tasks. To test our proposed approach, we have employed two different types of dataset: Dataset I measures the brain response to error and correct feedback, where the error feedback is assumed to evoke Error related Potential (ErrP) signals \cite{schie2004}, \cite{chav2014}. ErrP signals indicate awareness of the subject towards an occurrence of error. Dataset II comprises motor imagery signals \cite{pfurt1997},\cite{leeb2007} related to right hand and foot movement.

\section{Materials and Methodology}

\begin{figure}[!t]
\centering
\includegraphics[width=3.5in]{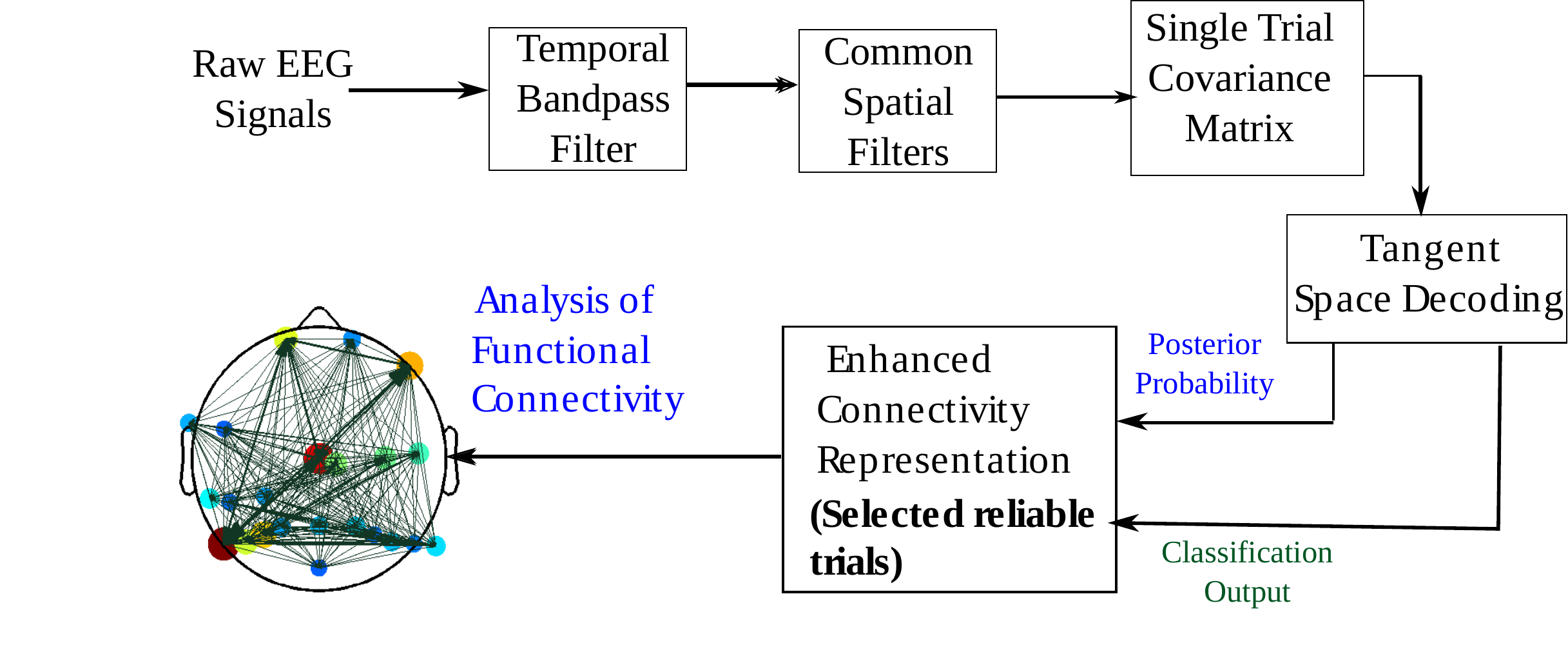}
\caption{A simplified block diagram of the proposed method for dual purpose of error potential decoding and connectivity measure enhancement.}
\label{fig_scheme}
\end{figure}

In this paper, we enhance the functional connectivity analysis of right hand and foot motor imagery (MI) signals and ErrP signals by removing the irrelevant trials from the relevant ones. The adjacency matrix of the graph is derived from the  covariances of the different tasks. The covariance matrices reflects the pairwise interaction across multiple channels and it contains the spatial information embedded in an EEG signal \cite{far2009}, which is employed for functional connectivity analysis. Then, we have designed a tangent space logistic regression classifier, which is similar to the classifier designed in \cite{barac2014} and is capable of detecting an occurrence of MI and ErrP across different sessions for the same subject. 

The step-by-step process involved in our proposed approach is as follows: First, we determine the interaction among the selected set of optimal channels from the covariances of each trial using Common Spatial Patterns (CSP) algorithm. Then, we use this covariance matrix as input to the designed Tangent space Logistic Regression classifier to detect the Error and noError trials for Dataset I and Right hand and foot imagery for Dataset II. Simultaneously, we also separate the relevant trials from the irrelevant ones by selecting those trials which have a posterior probability greater than 0.7. This allows us to select only trials of high confidence and thus, reduce the chance of including irrelevant trials, identified from their high chance of misclassification. Lastly, we perform graph theroretical analysis to study the interactions among the optimal set of selected channels. Fig. \ref{fig_scheme} illustrates a simplified block diagram of our proposed approach. 

\subsection{Dataset Description}

\begin{figure}[t]
\centering
\includegraphics[width=3.5in]{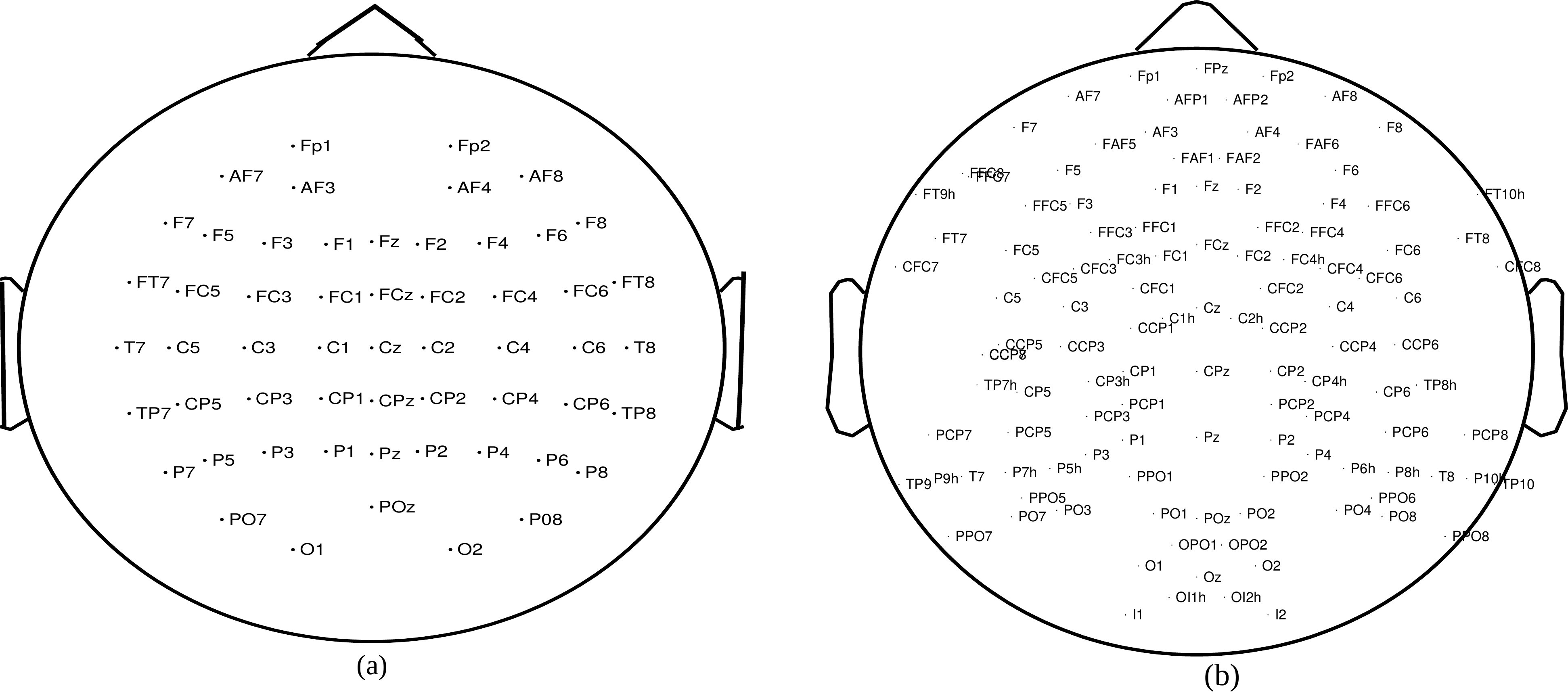}
\caption{Electrode locations of the 56 channels in (a) Dataset I (b) Dataset II arranged in an extended 10-20 system.}
\label{fig_elec}
\end{figure}

\subsubsection{Dataset I}

Dataset I is obtained from the `BCI Challenge @ NER 2015' competition hosted at Kaggle \cite{elec1}. The EEG dataset contains recording from 26 participants in the age range of 20-37 years and none of the subjects had any previous experience with any BCI application. Prior to the experiment, the participants signed an informed consent approved by the Local Ethical Committee \cite{elec1}.

The brain activities of the participants are recorded with 56 passive Ag/AgCl EEG sensors (VSM-CTF compatible system) and the placement of the electrodes followed the extended 10-20 system (Fig.\ref{fig_elec}(a)). The electrodes were referenced to the nose, the ground electrode was placed on the shoulder and the impedances of the electrodes were maintained at 10 k$\Omega$. During acquisition, the signals are sampled at 600 Hz but to aid in online processing, the signals provided are down-sampled to 200 Hz.

Each participant underwent 5 separate sessions of copying the spelling of a 5-letter word using P300 which is discussed in \cite{perrin2012}. The first four sessions are made of 12 five-letter words and the fifth (last) session comprises 20 five-letter words. So, for the first four sessions there are 60 feedback periods and for the fifth session there are 100 feedback periods. For this study, we have trained our decoder using the first four sessions for the same subject and tested it on the fifth session. The class labels for all the sessions are provided by the organizers.

Each trial comprises a feedback period where the decoded output is displayed on screen to the subject. If the decoded output doesn't match that of the target letter, than the subject recognizes an error has occured and thus, elicits an ErrP. Following the feedback session, a 0.5 second break is incorporated which marks the end of the current trial \cite{perrin2012}. For this paper, we are only working with the feedback portion (1.3 seconds) of each trial for each participants to detect ErrP potential in the EEG. We name the feedback trials containing ErrP as '\emph{Error}' and the trials with no ErrP as '\emph{noError}'.

\subsubsection{Dataset II}

We have employed Dataset IVa on motor imagery task of the BCI Competition III (http://www.bbci.de/competition/iii/desc\_IVa.html) as Dataset II for this study. This dataset was prepared from the recordings of five subjects, identified as, \emph{aa}, \emph{al}, \emph{av}, \emph{aw} and \emph{ay}. The EEG was recorded using BrainAmp amplifiers and a 128 channel Ag/AgCl electrode cap from ECI, of which 118 channels (Fig.\ref{fig_elec}(b)) were used for measurement arranged in the extended 10-20 system. The signal were band-pass filtered between 0.05 and 200 Hz and sampled at 1000Hz. Finally, the signal is downsampled at 100Hz for analysis. Each subject has undergone 280 trials of 3.5 seconds for the following two motor imagery tasks: right hand and right foot. In our study, we have employed the first 200 trials for training and the remaining 80 trials for validation. The class labels were provided for all the trials.  
 
\subsection{Pre-processing}

Along with the relevant EEG corresponding to the brain activity of the task performed by the participant, the signals acquired from the EEG recorder may also consist of information acquired from other brain activities (not related to the tasks). This EEG termed as \emph{background EEG}, may be detrimental to the detection of the signature features of a particular task and thus, can be considered as noise. Other forms of noise prevalent in EEG occurs due to muscle or eye movement, line noise and other stray noise from the environment. To remove the artifacts and extract the relevant information from the signal, researchers employ different types of spatial or temporal filtering techniques \cite{dorn2007}.

\emph{Dataset I:} It is known from previous literature that ErrP signals are dominant in the frequency range of 0.1 to 10 Hz \cite{combaz2012},\cite{ferrez2008}. The incoming EEG signals (for each electrode channel) are band-pass filtered with a 0.1 to 10 Hz pass-band using an IIR (impulse invariant response) Butterworth filter \cite{opp1999} of order 5. The pass-band and stop-band attenuation are 1 and 50 dB, respectively. IIR filters are very efficient tools of filtering of digital signals and they require less computational time when compared to other filters \cite{opp1999}. After filtering, 1 second of EEG is extracted from the onset of every feedback period at each trial for further analysis.

\emph{Dataset II:} Motor Imagery signals, defined by \emph{Event Related Desyhcronization/ Synchronization}, are pre-dominant in the $\mu-$band (8-12Hz) and the central $\beta-$band (16-24 Hz) \cite{pfurt1997}, \cite{bhatt2016}. Thus, the raw EEG signals are band-pass filtered with a [8,12]Hz and [16,24]Hz bandwidth by employing an elliptical filter of 6th order \cite{opp1999}. Then, from the filtered data, epochs of 3.5 seconds (corresponding to the onset of visual cue) were extracted for further analysis.

\subsection{Covariance matrix construction using CSP algorithm}

Common Spatial Patterns (CSP), first used in EEG classification of motor imagery tasks by Ramoser \cite{ram2000}, are used to project the multi-channel EEG data into low-dimensional spatial subspace with a projection matrix, capable of maximizing the variance ratio of the two class signal matrices. It is based on the simultaneous diagonalization of the covariance matrices of both classes.

Let us define a single trial EEG as $\mathbf{X}\;\epsilon\;\mathbf{R}^{Ch \times T}$, where $Ch$ is the number of channels and $T$ is the number of samples per channel. Thus, by using the CSP algorithm one can project $\mathbf{X}$ to a spatially filtered counterpart $\mathbf{Z}$, by

\begin{equation}
\mathbf{Z} = \mathbf{WX}
\label{eq1}
\end{equation}

where, rows of the projection matrix $\mathbf{W}$ are the spatial filters and the columns of $\mathbf{W^{-1}}$ represents the CSP. The computation of the CSP is explained thoroughly in \cite{ram2000},\cite{wang2005}. The patterns obtained from the columns of $\mathbf{W^{-1}}$ can be considered as EEG topographies. Thus, the first and last few columns of the spatial patterns corresponds to the largest variance of one task and the smallest variance of the other. Thus, channels which correspond to the maximal coefficients of spatial pattern vectors are the channels most related to the task specific sources. This principle was first formulated by Wang et al. in \cite{wang2005}. In our study, we applied this principle to select six pairs of optimal channels (corresponding to the MI and ErrP activity) for graph visualization purposes. 

For classification and functional graph analysis, we then calculate the trial covariance matrices of the projected spatially filtered signals $\textbf{Z}_i$, where $i$ is the trial number, by using the sample covariance estimator:

\begin{equation}
{\mathbf{\Sigma}}_i = \frac{1}{N-1}{\mathbf{Z}}_i{\mathbf{Z}}_i^T.
\label{eq2}
\end{equation}


\subsection{Tangent Space Logistic Regression Classifier (TSLR)}

Covariance matrices, including the one in (\ref{eq2}), are symmetric positive definite (SPD) matrices \cite{jaya2013},\cite{barac2014} embedded into the Euclidean space \cite{yger2013}, \cite{yger2015}. It is reported in \cite{barac2013} that the space of SPD $E \times E$ matrices $P(E)$ forms a differentiable manifold $M$. \emph{Tangent space mapping}, is a powerful Riemannian geometry concept, which projects the covariance matrices as Euclidean objects into a lower dimensional space \cite{barac2014}. This mapping allows us to employ state-of-the-art classifiers using Riemannian framework to improve the performance. The Euclidean geometry suffers from disadvantages, such as, the SPD matrix space produces a non-complete space while using it and a swelling effect, explained in details in \cite{vinc2007}. To avoid these problems, we have used \emph{LogEuclidean metric} \cite{yger2013} in place of Euclidean metric.

For two covariance matrix, say $\tilde{\mathbf{\Sigma}}_1$ and $\tilde{\mathbf{\Sigma}}_2$, the \emph{LogEuclidean distance} $\delta_L$ is given by
\begin{equation}
\delta_L(\tilde{\mathbf{\Sigma}}_1,\tilde{\mathbf{\Sigma}}_2) = ||\text{log}(\tilde{\mathbf{\Sigma}}_1)- \text{log}(\tilde{\mathbf{\Sigma}}_2)||_F
\label{eq12}
\end{equation}

where $\text{log(.)}$ represents the matrix logarithm and $||.||$ stands for Frobenius norm.

Next, we calculate the \emph{logEuclidean mean} $\wp$ using $\delta_L$ for $I$ covariance matrices, as
\begin{eqnarray}
\wp(\tilde{\mathbf{\Sigma}}_i,\ldots,\tilde{\mathbf{\Sigma}}_I) =& \text{arg}\text{min} \sum_{i=1}^{I}\delta_L^2(\tilde{\mathbf{\Sigma}}, \tilde{\mathbf{\Sigma}}_i) \nonumber \\
=& \exp{(\frac{1}{I} \sum_i \log{\tilde{\mathbf{\Sigma}}_i})}
\label{eq13}
\end{eqnarray}

Next, we project the matrices from the manifold in a vector space called \emph{Tangent space} and the procedure is called \emph{tangent space mapping}. Each $E \times E$ covariance (SPD) matrix is represented by vectors of dimension $n(n+1)/2$ in the tangent space, given by

\begin{equation}
s_i = \text{log}[\text{upper}(\tilde{\mathbf{\Sigma}}_\wp^{-1/2}\tilde{\mathbf{\Sigma}}_i\tilde{\mathbf{\Sigma}}_\wp^{-1/2})].
\label{eq14}
\end{equation}
  
Now, the projections of the tangent space can be applied with any available standard classification algorithm for decoding. In this study, we have used L1 regularized Logistic Regression classifier \cite{alp2004}, \cite{hast2001} to detect ErrP from the EEG dataset. Logistic regression is a probabilistic type of classifier which predicts the outcome (or classes) of one or more features based on a logistic function, $g(z)=\frac{1}{1+e^{-z}}$ where $z$ is the linear combination of the input features. This classifier measures the relationship between the classes and the features by using the probability scores as the predicted value of the classes. In short, it predicts the probability of the class to be positive \cite{alp2004}, \cite{hast2001}. 

\subsection{Graph Analysis}
A graph is defined as a set of ordered pair $G = (V,E)$, where $V$ is a set of nodes and $E$ is a set of edges associated with two nodes. In the present context, the selected channels are considered as nodes and the connection weights between each electrodes represent the edges. $\Sigma_i$ from Eq.(\ref{eq2}) is used to construct the connection matrix \cite{fall2010} which determines the pairwise interaction among different channels. Various measures exist in literature \cite{bull2009} to study the connections in a brain network but in this study we have employed four such measures.

\begin{enumerate}

\item {\bf Local Clustering Coefficient:} Clustering coefficient is a measure of how nodes in a graph tends to cluster together. The local clustering coefficient of a node is a measure of how close its neighbors are from being a complete graph. Neighborhood of a node V is the subgraph of vertices connected by an edge E to V.  

\item{\bf Participation Coefficient:} It is a measure of diversity of intermodular connections of individual nodes. Modularity is a measure that determines the strength of division of a graph into modules and the links between different modules are known as intermodular connections.   

\item{\bf Local Efficiency:} It is the average of inverse shortest path computed on the neighborhood of the node.

\item {\bf Node Strength:} Node strength \cite{bull2009} for a weighted graph is calculated as the sum of the outgoing intensity from a node and the incoming intensity to it from other nodes. It shows the amount of connections each node has with other nodes in the graph \cite{ghosh2015}.

\end{enumerate}

\section{Results and Discussion}

The analysis of the work has been done on a Python environment in an ubuntu 15.10 based computer system with 8GB ram and AMD A10 1.89 GHz processor. 

\subsection{Classification Results} 

We evaluate the performance of the TSLR decoder from the following performance metrics: Accuracy, Precision and Recall \cite{alp2004}. For Dataset I, we have employed the first 4 sessions to train the decoder and the fifth session to test it. For Dataset II, the first 200 trials were used for calibrating the decoder and the remaining 80 trials for testing.

\begin{figure}[!t]
\centering
\includegraphics[width=3in]{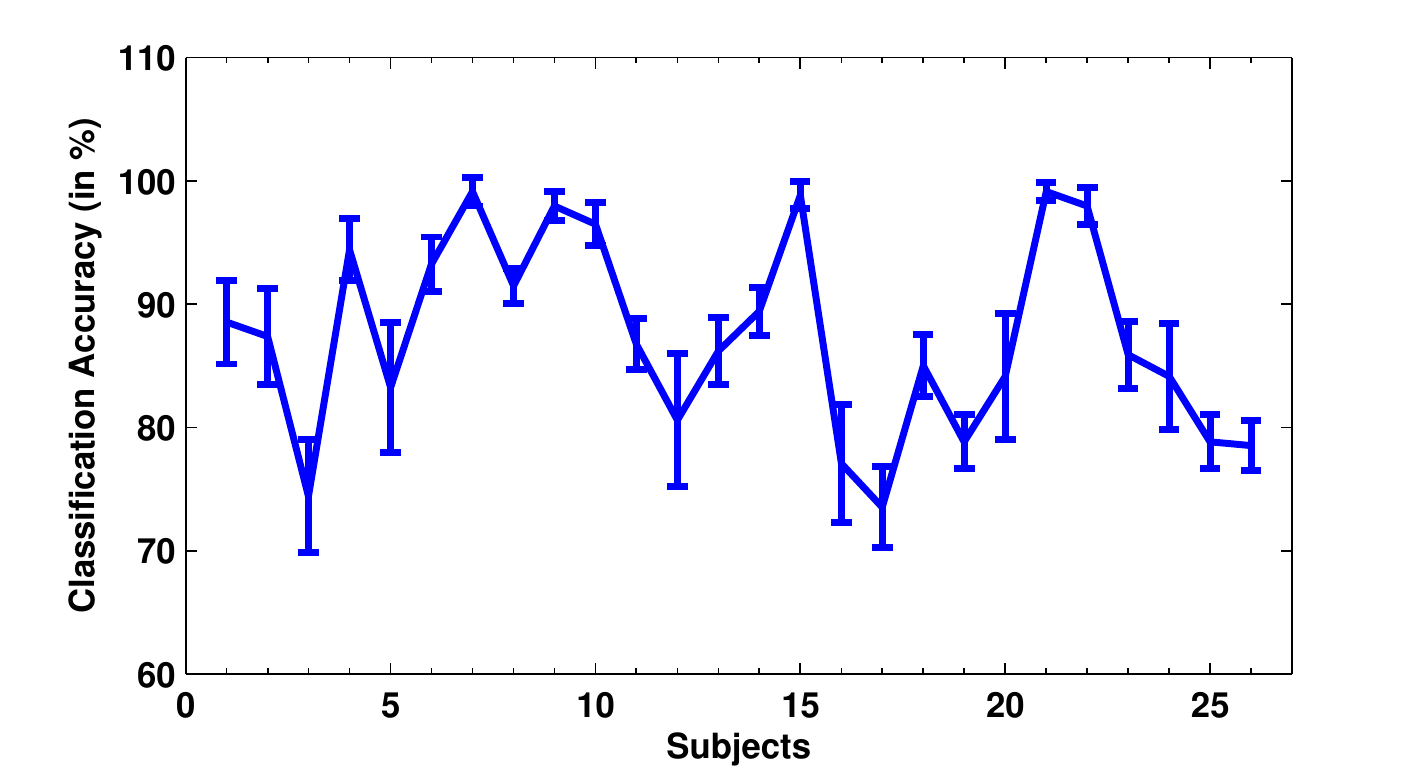}
\caption{Mean accuracy and standard deviation of the TSLR classifier for 26 subjects after 10-fold cross validation.}
\label{fig_cv_err}
\end{figure}

\begin{figure}[!t]
\centering
\includegraphics[width=3in]{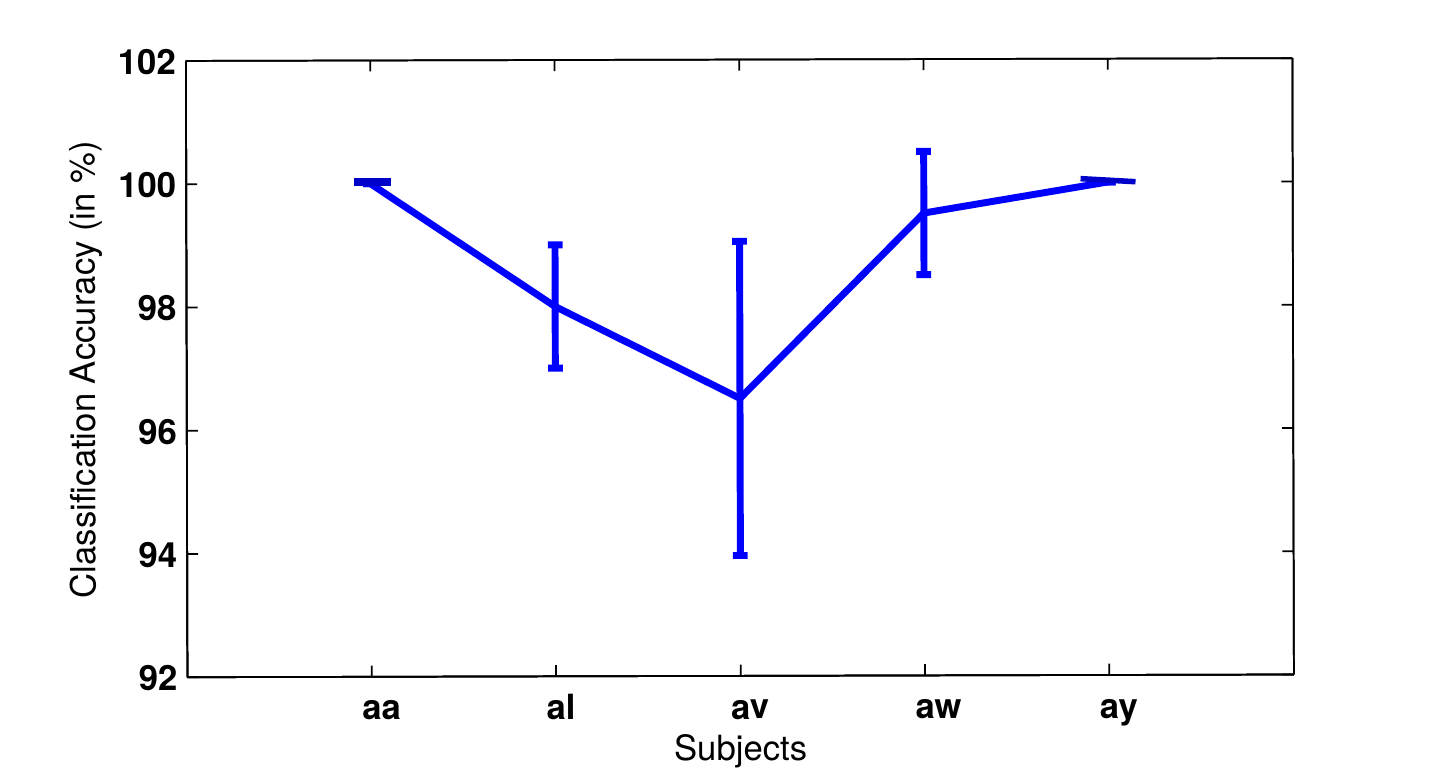}
\caption{Mean accuracy and standard deviation of the TSLR classifier for 5 subjects after 10-fold cross validation.}
\label{fig_cv_mi}
\end{figure}

To analyze the performance of the decoder during calibration, we have used $k$-fold cross validation technique \cite{alp2004}, where we have taken the value of $k$ to be 10 to lower the variance of the outcome. The average classification accuracy and its standard deviation for each subject are shown in Fig. \ref{fig_cv_err} for Dataset I and Fig. \ref{fig_cv_mi} for Dataset II. From the Fig. \ref{fig_cv_err} it is observed that Subject 21 has the highest classification accuracy of 99.12\% $\pm$ 0.72, while Subject 17 has the lowest classification accuracy of 73.52\% $\pm$ 3.27. The mean accuracy for all subjects of Dataset I is 87.34\% $\pm$ 8.01 and 20 of the 26 subjects shows an accuracy over 80\%. From the Fig. \ref{fig_cv_mi} it is observed that Subject `\emph{aa}' and `\emph{ay}' has the highest classification accuracy of 100\%, while Subject `\emph{av}' has the lowest classification accuracy of 96.50\% $\pm$ 2.55. The mean accuracy for all subjects of Dataset II is 98.80\% $\pm$ 1.52.    

Next, we test the training of the TSLR classifier for each dataset. Table \ref{table_train_err} presents classification accuracy, precision and recall of the 26 subjects of Dataset I and Table \ref{table_train_mi} presents the same for the 5 subjects of Dataset II. From Table \ref{table_train_err}, it is observed that the accuracy ranges from 67\% to 100\% with mean accuracy, precision and recall of 84.35\%, 88.30\% and 82.71\%, respectively. The results of most subjects are comparable to their training score with the average difference between the calibrated and the validated results being 2.99\%. The highest difference in performance is noted in Subject 12 from 80.58\% during training to 70\% during validation (difference of 10.58\%). It is also noted that for most subjects the precision and recall is over 80\% which suggest intuitively that the classifier can detect over 80\% of the positive samples (in our study, error feedbacks) without misclassifying it. 

From Table \ref{table_train_mi}, it is observed that the accuracy ranges from 77.10\% to 100\% with mean accuracy, precision and recall of 89.67\%, 88.69\% and 91.67\%, respectively. Subject `\emph{aa}' and `\emph{av}' have a huge difference of 16.25\% and 19.40\% in between the training and validated results. The high variability of the results may be attributed to the training of the subject and the decoder. The results of the other 3 subjects are comparable to their training score with the average difference between the calibrated and the validated results being 3.33\%.  From the precision results of Dataset II, Subject `\emph{aa}' and `\emph{av}' are weak in correctly classfying the positive class, which in this study, is the right motor imagery. 
 
\begin{table}[!t]
\caption{Classification Results of the 5th session data (100 trials) for all subjects}
\label{table_train_err}
\centering
\begin{tabular}{|c|c|c|c|}
\hline
{\bf Subject} & {\bf Accuracy}	& {\bf Precision}	& {\bf Recall} \\
{\bf ID} & {in \%}	& {in \%}	& {in \%} \\
\hline
01 & 84.00 & 92.10 & 87.50 \\
02 & 87.00 & 85.25 & 92.86 \\
03 & 75.00 & 65.71 & 63.89 \\
04 & 89.00 & 95.52 & 88.89 \\
05 & 85.00 & 83.33 & 64.51 \\
06 & 92.00 & 94.50 & 96.63 \\
07 & 99.00 & 98.88 & 100.00 \\
08 & 90.00 & 89.86 & 98.38 \\
09 & 93.00 & 91.86 & 100.00 \\
10 & 95.00 & 95.65 & 98.86 \\
11 & 84.00 & 83.33 & 89.29 \\
12 & 70.00 & 82.35 & 64.14 \\
13 & 92.00 & 93.75 & 90.00 \\
14 & 85.00 & 92.68 & 76.00 \\
15 & 100.00 & 100.00 & 100.00 \\
16 & 67.00 & 76.32 & 54.72 \\
17 & 71.00 & 73.16 & 81.36 \\
18 & 81.00 & 82.14 & 94.52 \\
19 & 73.00 & 93.54 & 53.70 \\
20 & 78.00 & 86.21 & 78.12 \\
21 & 93.00 & 92.39 & 100.00 \\
22 & 97.00 & 96.81 & 100.00 \\
23 & 84.00 & 92.45 & 80.33 \\
24 & 75.00 & 78.26 & 70.58 \\
25 & 73.00 & 93.55 & 63.70 \\
26 & 81.00 & 86.20 & 62.50 \\
\hline
Mean & 84.35 & 88.30 & 82.71 \\
\hline
\end{tabular}
\end{table}

\begin{table}[!t]
\caption{Classification Results of the 5th session data (100 trials) for all subjects}
\label{table_train_mi}
\centering
\begin{tabular}{|c|c|c|c|}
\hline
{\bf Subject} & {\bf Accuracy}	& {\bf Precision}	& {\bf Recall}\\
{\bf ID} & {in \%}	& {in \%}	& {in \%}\\
\hline
aa & 83.75 & 77.92 & 90.00 \\
al & 100.00 & 100.00 & 100.00 \\
av & 77.10 & 72.96 & 85.00 \\
aw & 93.75 & 97.44 & 90.48 \\
ay & 93.75 & 95.12 & 92.85 \\
\hline
Mean & 89.67 & 88.69 & 91.67 \\
\hline
\end{tabular}
\end{table}

The high classification results for each subject across different sessions suggests that the proposed TSLR classifier is robust and efficient in detecting ErrP and MI signals. By implementing an adaptive process in tuning the parameters to each subject, the performance of the classifier can be made more robust and show a less variable accuracy across different subjects. The trials which were correctly classified and had a posterior probability $\geq$ 0.7 (relevant trials) in the validation stage are then extracted for graph analysis.

\subsection{Average channel interaction of the projected signal}

\begin{figure}[!t]
\centering
\includegraphics[width=3.5in]{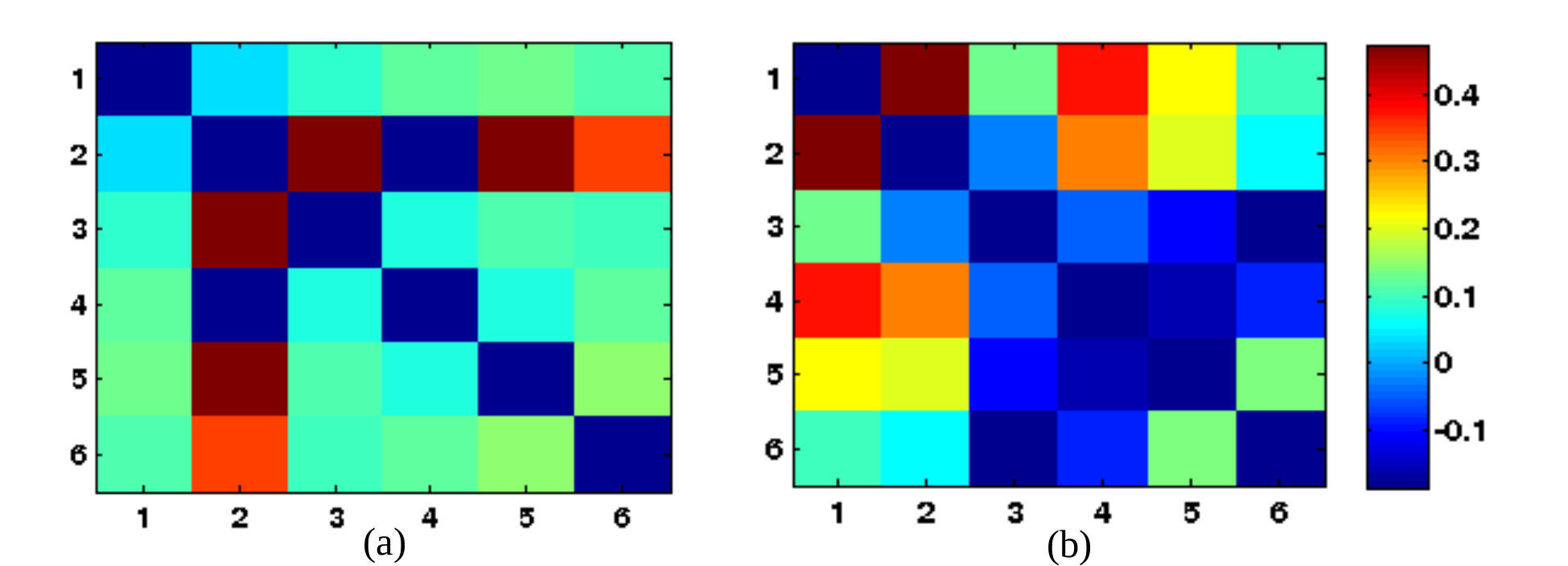}
\caption{Average covariances for the six spatially projected channels after CSP of 26 subjects of Dataset I, where (a) corresponds to `Error' trials (b) corresponds to `noError' trials.}
\label{fig_err_cov}
\end{figure}

\begin{figure}[!t]
\centering
\includegraphics[width=3.5in]{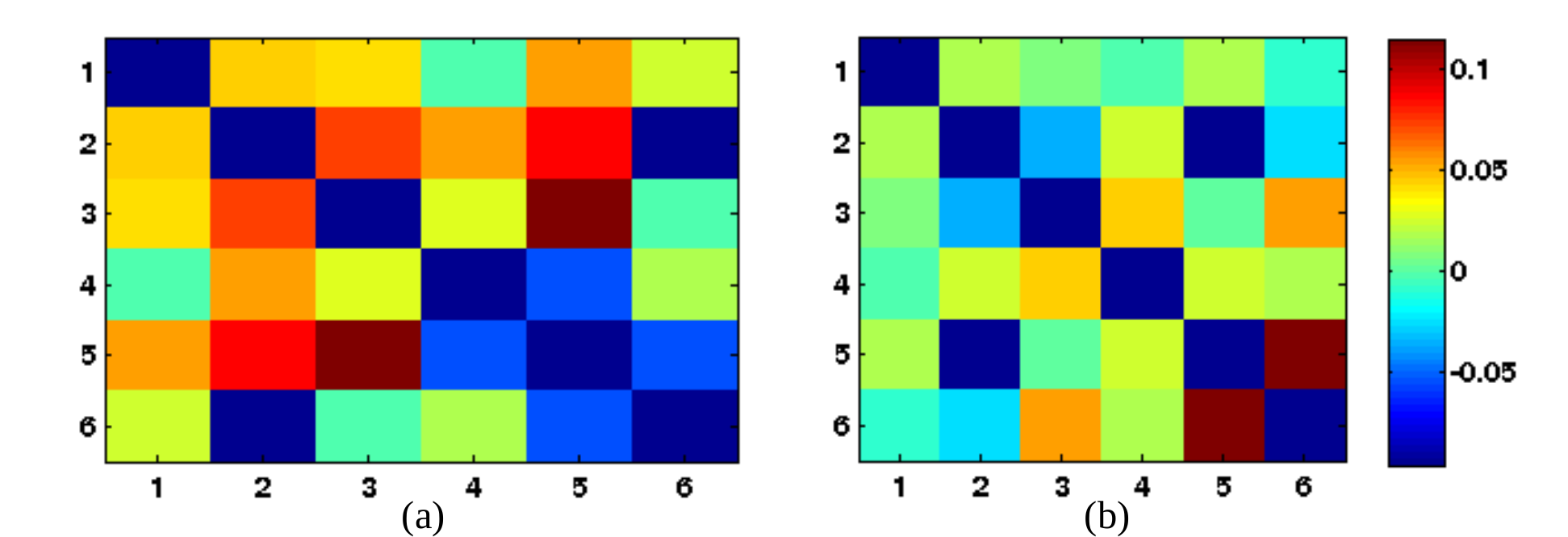}
\caption{Average covariances for the six spatially projected channels after CSP of 5 subjects of Dataset II, where (a) corresponds to right motor imagery (b) corresponds to foot motor imagery.}
\label{fig_bci4a_cov}
\end{figure}

After filtering and epoching the raw EEG signals, we calculate the covariances between the six channels for the CSP projected signal. We show the average channel interaction through the covariances of the projected signals for all subjects of dataset I and II in Fig. \ref{fig_err_cov} and Fig. \ref{fig_bci4a_cov}, respectively. Fig. \ref{fig_err_cov}(a) shows the pairwise interaction for `\emph{Error}' trials, channel2 shows maximum interaction with channel3 and channel5. Fig. \ref{fig_err_cov}(b) displays the same for `\emph{noError}' trials where the highest pairwise interaction is between channel1 and channel2. Fig. \ref{fig_bci4a_cov}(a) and (b) shows the pairwise interaction for `right' and `foot' motor imagery, respectively. For right imagery, it is noted that channel3 has the highest with channell5, followed by interaction of channel2 with channel5 and channel3. On the other hand, for foot imagery, channel5 shows the maximum interaction with channel6. As observed from the figures, the pairwise interaction for different tasks are widely different from each other, which indicates that different processes of the brain are active during separate tasks. 

\subsection{Connectivity Analysis on the selected trials}

We select the relevant trials after the classification step, average them, and then perform reliable functional connectivity analysis. Here, we have used clustering coefficient, local efficiency, participation coefficient and node strength to determine the graph properties but other network measures can also be used. 

\begin{figure}[!t]
\centering
\includegraphics[width=3.5in]{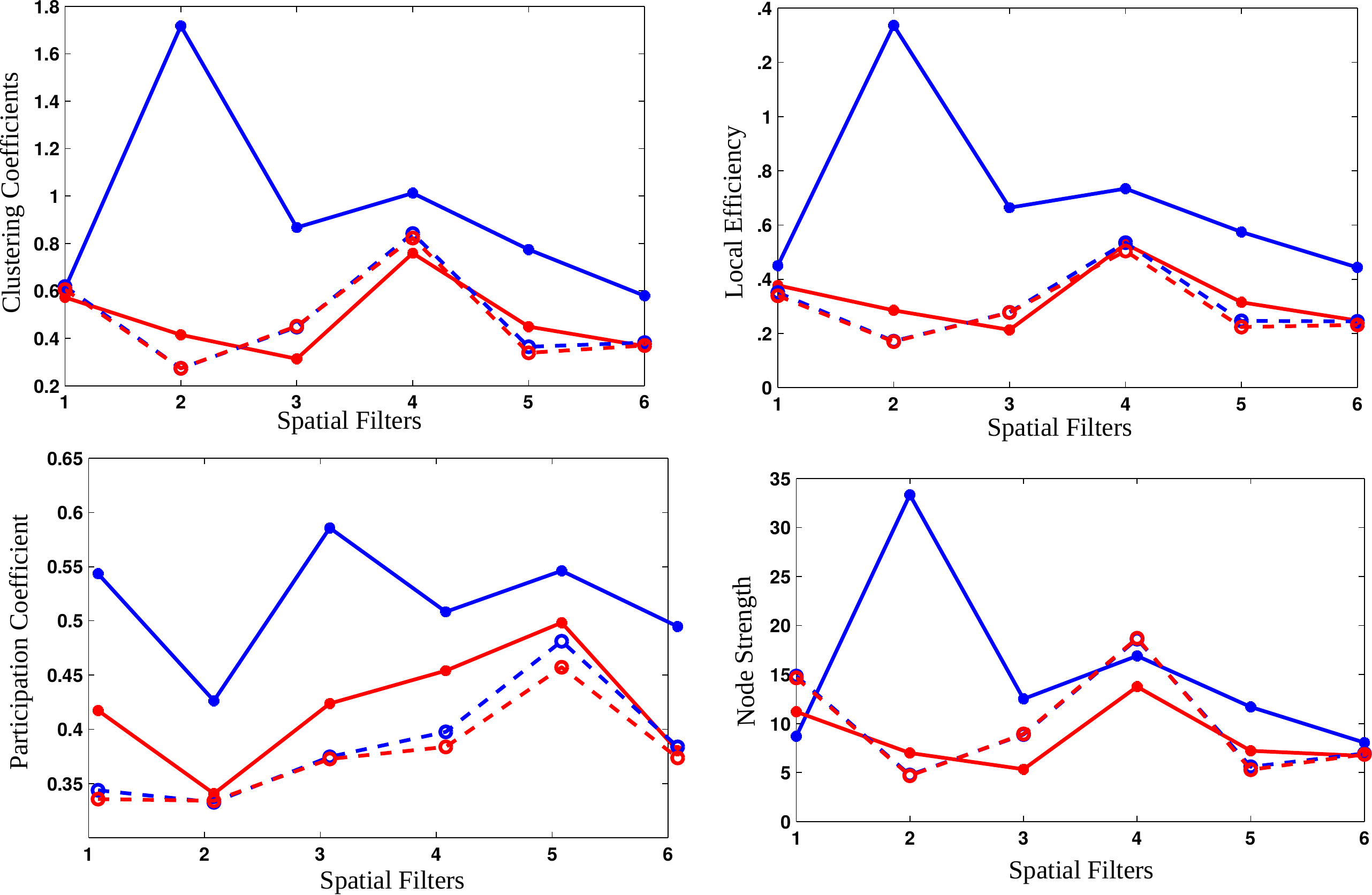}
\caption{The average clustering coefficient, local efficiency, participation coefficient and node strength of each projected channels for dataset I. The blue and red color indicates `Error' and `noError' trials, respectively and the solid and dashed lines indicates selected relevant trials and all trials, respectively.}
\label{fig_graph_err}
\end{figure}

\begin{figure}[!t]
\centering
\includegraphics[width=3.5in]{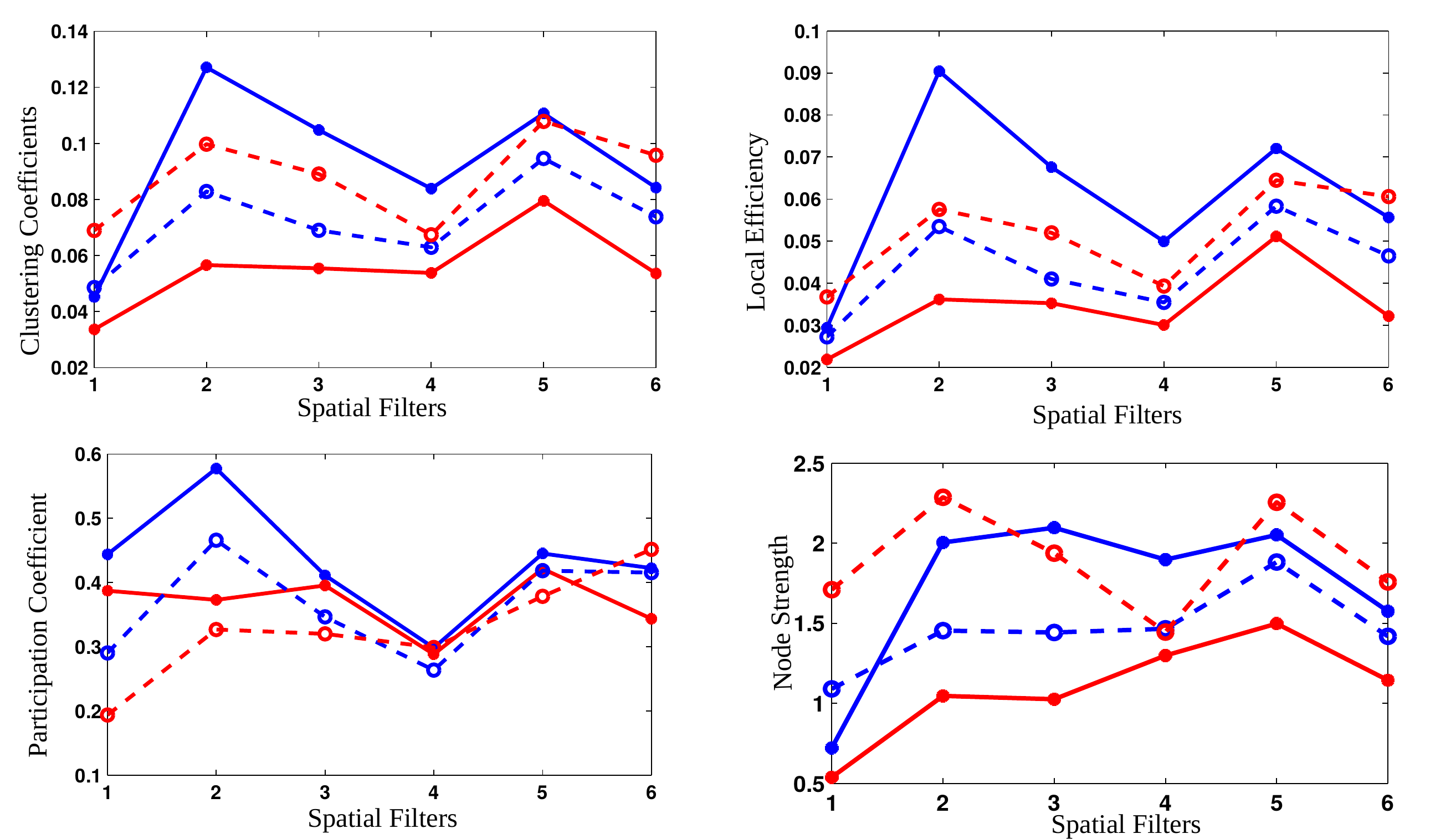}
\caption{The average clustering coefficient, local efficiency, participation coefficient and node strength of each projected channels for dataset II. The blue and red color indicates right and foot motor imagery trials, respectively and the solid and dashed lines indicates selected relevant trials and all trials, respectively.}
\label{fig_graph_bci4a}
\end{figure}

Fig.\ref{fig_graph_err} and \ref{fig_graph_bci4a} provides the average values across all subjects of each graph parameters for dataset I and II. As noted from both the figures, all parameter values for the complete trials (shown by dotted lines) are closely spaced to each other while the same for the selected trials (shown by solid lines) are more widely spaced. For example, for dataset I, the participation coefficients for all the `Error' and `NoError' trials are close to each other that it would be difficult to differentiate between the two while observing the graph parameters but in selecting only the correctly classified trials we can observe the differentiability between the two classes increase and so does the visualization of the graph parameters along with it.   

\begin{figure}[!t]
\centering
\includegraphics[width=3.5in]{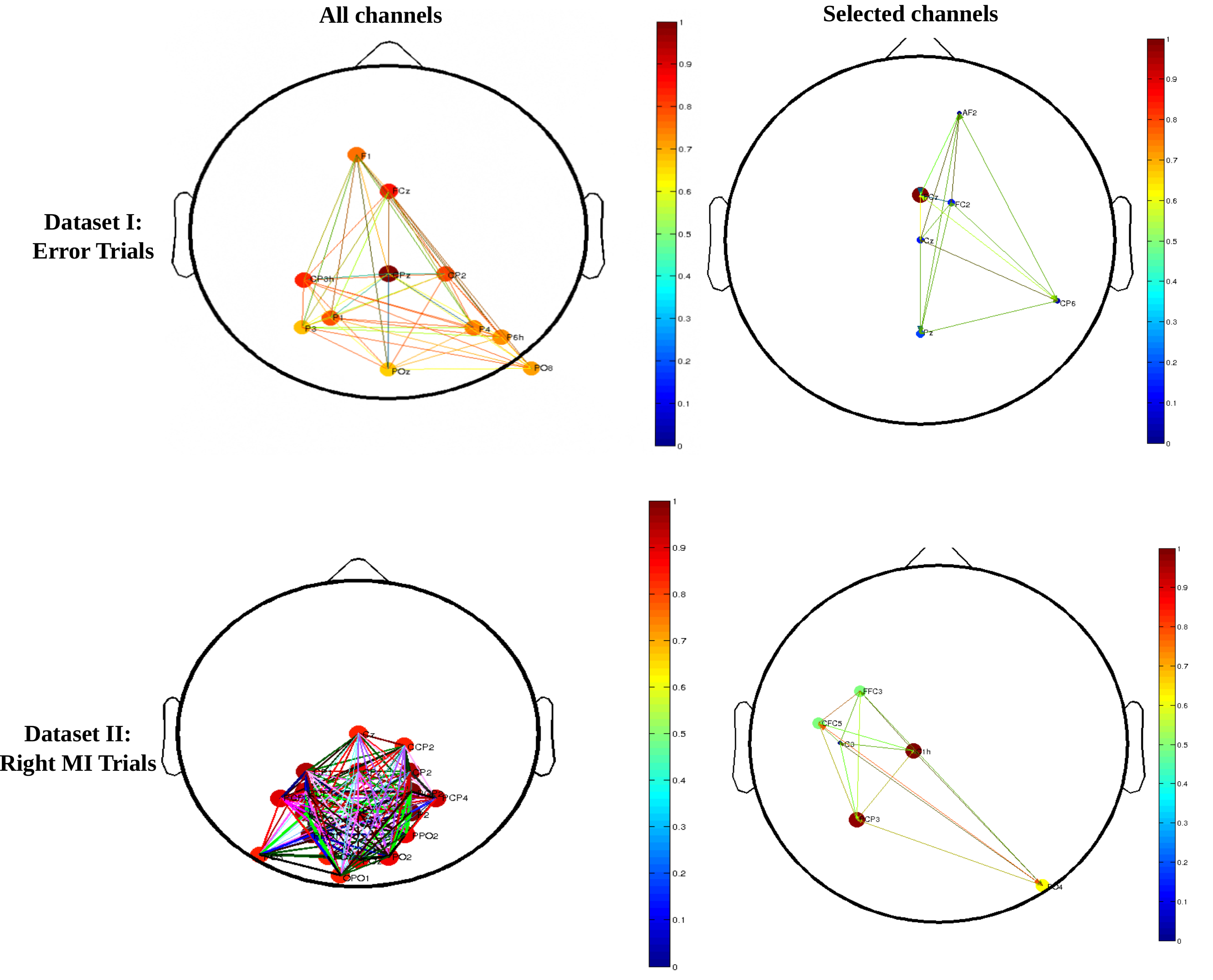}
\caption{An example of graph visualization for all channels and the selected relevant channels. The error trials from subject 13 dataset I and the right MI trials from subject `ay' of dataset II are taken in this example. The dots represents the node strengths of each channels and the lines represents the edge weights connecting two channels. The value of the node strengths and edge weights are represented by the color bar.}
\label{fig_conn_examples}
\end{figure}

Fig.\ref{fig_conn_examples} gives an example of graph visualization by using the node strengths of subject 13 of dataset I and subject `\emph{ay}' of dataset II. It must be noted that for the plots of all channels, the figure is constructed for the best 10\% channels of node strength. For the error trials in dataset I, the selected channels \cite{wang2005} are located on the midline, i.e., FCz, Cz and Pz, but while including all channels for analysis, major brain activities are shown to concentrate in the parietal and occipital region. It is known from literature that ErrP signals originate in the mid-anterior cingulate cortex which corresponds well to the selected channels. The high strength of FCz, which lies over the anterior cingulate cortex, correctly confirms the observance of error by the subject. For the right MI trial of dataset II, the selected channels are concentrated at the left side of the brain (contralateral to right MI) with the strongest channels being C1 and CP3 and C5 being the weakest channels. Thus the contralateral side of the brain is active for the right motor imagery task. By selecting the reliable brain signals specifically relevant to the targeted task, it could be possible to remove the non-relevant mental states and to highlight the targeted functional connectivity which is really necessary to be extracted to analyze the brain signals in a reliable manner. Thus, through this example we show the advantage of our proposed connectivity analysis approach. 

\section{Conclusion}

In this study, we have proposed a novel technique to enhance functional connectivity analysis in a systematic way by removing badly classified (irrelevant) trials from datasets consisting of error related potential and motor imagery EEG signals. To improve graph analysis we have also implemented an approach of electrode channel selection, where the functional connectivity is applied on optimal subset of electrodes. Thus, at first, we have calculated the covariances of Right Hand MI, Foot MI and ErrP from their respective EEG signals. Then, we decoded every trials using tangent space logistic regression classifier (with the covariances as input) to select only the correctly classified trials with a posterior probability $\geq$ 0.7, which were then used for functional connectivity analysis. As noted from the results, by selecting only the relevant trials, the separability among the graph parameters for two tasks have greatly improved. In most BCI protocols, the experimental brain signal data can be easily influenced by other cognitive tasks which can occur in parallel to the targeted cognitive tasks. This fact can affect the quality of functional connectivity analysis as non-related cognitive state can hide the brain network reaction to the targeted task. Thus, the proposed approach in this paper helps in improving the brain connectivity to its associated cognitive tasks. This approach to the best of the authors' knowledge is a first of its kind in enhancing brain connectivity studies in a systematic manner over different mental tasks (error and motor imagery). With modifications in the classification parameter, this method can be implemented for other BCI protocols and even for general multi-channel brain signal analysis to extract reliable sessions systematically. This approach would contribute to the general functional connectivity extraction for the targeted cognitive states. 

\bibliographystyle{unsrt}  

%
%
%
%

\bibliographystyle{IEEEtran}
\bibliography{draft}

\end{document}